\documentclass[aps,twocolumn,amsmath,amssymb,showkeys,showpacs,footinbib,superscriptaddress]{revtex4-1}
\usepackage{epsfig,graphicx}
\usepackage{setspace}
\usepackage[english]{babel}
\usepackage{amsfonts}
\usepackage{amsmath}
\usepackage{latexsym}
\usepackage{graphics,bm}
\usepackage{natbib}
\usepackage{dcolumn}
\usepackage{bm}
\usepackage{rotating}
\usepackage{amsmath}
\usepackage{bbold}
\usepackage{braket}
\usepackage{epstopdf}
\usepackage{color}

\usepackage{soul,xcolor}
\setstcolor{red}
\usepackage[T1]{fontenc}
\usepackage[applemac]{inputenc}
\usepackage{lmodern}
\usepackage{ae}
\usepackage{units}
\usepackage{color}
\usepackage{url}
\usepackage[colorlinks]{hyperref}
\hypersetup{%
        plainpages=true,
        breaklinks=true,
        hypertexnames=false,
        pageanchor=true,
        colorlinks=true,
        linkcolor={blue},
        citecolor={red},
        urlcolor={blue},
        anchorcolor={black}
      }
\newcommand{\be}{\begin{equation}}
\newcommand{\ee}{\end{equation}}
\newcommand{\ba}{\begin{array}{lll}}
\newcommand{\ea}{\end{array}}

\newcommand{\lan}{\langle}
\newcommand{\ran}{\rangle}

\begin{document}
\title{Lasing in the superradiant crossover regime}
\author{Kamanasish Debnath}
\affiliation{Department of Physics and Astronomy, Aarhus University, Ny Munkegade 120, DK-8000, Aarhus C, Denmark}
\author{Yuan Zhang}
\email[e-mail:]{yzhang@phys.au.dk}
\affiliation{Department of Physics and Astronomy, Aarhus University, Ny Munkegade 120, DK-8000, Aarhus C, Denmark}
\author{Klaus M{\o}lmer}
\affiliation{Department of Physics and Astronomy, Aarhus University, Ny Munkegade 120, DK-8000, Aarhus C, Denmark}
\begin{abstract}
A new class of laser, which harnesses coherence in both light and atoms, is possible with ultra-cold alkaline earth atoms trapped in an optical lattice inside an optical cavity. Different lasing regimes, including superradiance, superradiant and conventional lasing, are distinguished by the relative coherence stored in the atoms and in the cavity mode. We analyze the physics in two different experimentally achievable regions of the superradiant lasing regime. Our calculations confirm the narrow linewidth of superradiant lasing for the doubly forbidden clock transition ${}^3 P_0 \to {}^1 S_0$  of  strontium-87 atoms. Under strong driving of the dipole-forbidden transition ${}^3 P_1 \to {}^1 S_0$ of strontium-88 atoms the superradiant linewidth narrows further due to the coherent excitation of the cavity field.
\end{abstract}


\maketitle

\section{Introduction}
\begin{figure}
\begin{centering}
\includegraphics[scale=0.45]{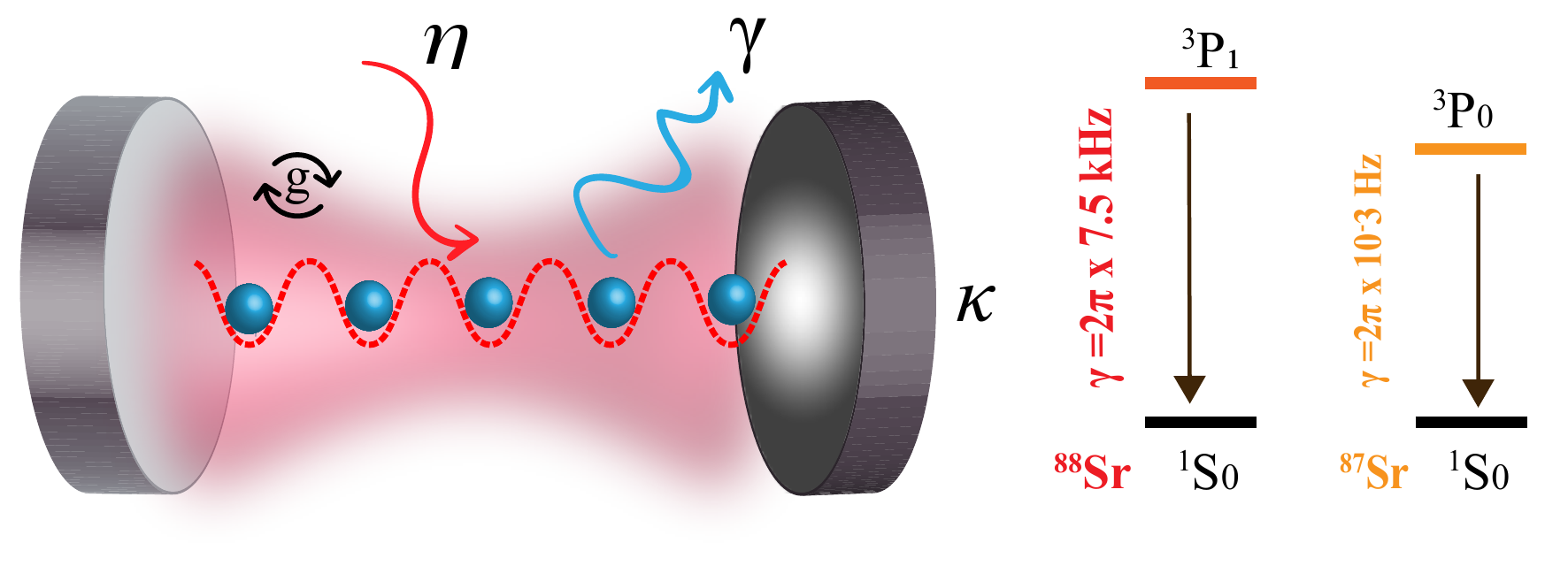}
\par\end{centering}
\caption{Many atoms are trapped in an one-dimensional optical lattice and coupled to the fundamental mode of an optical cavity with a strength $g$. The atoms decay and are incoherently pumped with rates $\gamma$ and $\eta$ while the cavity photons decay with a rate $\kappa$. We compare different regions in the superradiant lasing regime by studying the systems using the strontium-88 transition ${}^3 P_1 \to {}^1 S_0$~\cite{PhysRevX.6.011025} and the strontium-87 optical clock transition ${}^3 P_0 \to {}^1 S_0$~\cite{Norciae1601231}.}
\label{F1}
\end{figure}

In a conventional laser~\cite{Phy.Rev.99.1264}, amplification and optical phase coherence is established by stimulated photon emission from a population-inverted medium. This results in the Schawlow-Townes~\cite{PhysRev.112.1940} spectral line-width, inversely proportional to the photon number in the cavity. Seminal work by Dicke~\cite{PhysRev.93.99} suggested that the coherence can also be stored in the emitters that constitute the gain medium, if they interact collectively with common radiation field modes. Earlier studies concentrated on transient radiation caused by the collective superradiant decay of initially excited emitters. Recent theoretical~\cite{PhysRevLett.102.163601,PhysRevA.81.033847, PhysRevLett.113.154101} and experimental~\cite{Bohnet2012,PhysRevX.6.011025,Norciae1601231} studies showed that steady-state superradiance may yield lasing with millihertz line-width from ultra-cold alkaline earth atoms trapped in an optical lattice inside an optical cavity, see Fig.\ref{F1}.

Such a superradiant laser can operate either in a superradiant regime with less than one cavity photon and only atomic coherence ~\cite{PhysRev.93.99} or in a superradiant crossover regime with multiple photons and coherences in both the emitters and the cavity field ~\cite{TieriHolland}. In contrast, in the conventional lasing regime the atomic coherence does not play any significant role.

We employ second order mean-field theory~\cite{PhysRevLett.102.163601} to investigate two realistic systems with the aim of comparing the different physics in the atom-coherence dominated superradiant lasing regime. More precisely, we consider the dipole-forbidden transition ${}^3 P_1 \to {}^1 S_0$ with a decay rate $\gamma = 2\pi \times 7.5$  kHz  of bosonic strontium-88 atoms, which was used in~\cite{PhysRevX.6.011025} to demonstrate lasing in the deep superradiant lasing regime, and the doubly forbidden optical clock transition ${}^3 P_0 \to {}^1 S_0$ with a decay rate  $\gamma = 2\pi$ mHz  of fermionic strontium-87 atoms, which was used in~\cite{Norciae1601231} to show superradiance pulses. In these experiments, the fundamental cavity mode has a loss rate $\kappa = 2\pi \times 160$ kHz and couples resonantly to the ${}^3 P_1 \to {}^1 S_0$ transition with a strength $g = 2\pi \times 10.6$ kHz, and to the ${}^3 P_0 \to {}^1 S_0$ transition with a strength $g = 2\pi \times 2.41$ Hz.

Since the atoms are trapped along the cavity axis, they couple to the fundamental cavity mode with same strength and we shall assume identical coupling, decay, dephasing and incoherent excitation rate for all atoms. Under this assumption,  the system state is symmetric under permutation of the atoms and the collective behavior is well captured by second order moments~\cite{PhysRevLett.102.163601}, which constitute a small set of variables and are readily calculated. However, in order to visualize the emitter dynamics and interpret the results, we shall employ the intuitive picture of the Dicke states, characterized by angular momentum quantum numbers $J,M$~\cite{PhysRev.93.99}. To unravel the mechanism responsible for the spectrum line-width, we simulate the coupling of the system to a filter cavity.  These theoretical tools allow us to go beyond first order mean field theory. More importantly, we will examine the interplay between optical excitation, coherence and correlations in the system and their consequences for the emission linewidth.

This article is organized as follows. In the following section, we present the quantum master equation for our system. In Sec. \ref{Radiation} and \ref{Spectrum} we present the simulations for steady-state radiation and spectrum from the system and analyze the influence of the incoherent pumping and the atom number. In the end, we provide concluding remarks and comment on possible developments in future.  

\section{Quantum Master Equation}
We consider $N$ identical two-level atoms with a frequency $\omega_a$ and a decay rate $\gamma$, a dephasing rate $\chi$, as well as a pumping rate $\eta$.  $\eta$ may be realized by exciting the atoms from the lower level to a higher excited level from which they decay rapidly to the upper level of our lasing transition~\cite{PhysRevLett.102.163601}.  The atoms are coupled with a strength $g$ to an optical cavity mode with frequency $\omega_c$ and photon loss rate $\kappa$. This system is described by the quantum master equation ($\hbar=1$ throughout the paper):
\begin{eqnarray}
&& \frac{\partial{}}{\partial t} \rho = -i[H,\rho] + \kappa \mathcal{L}[a]\rho  + \gamma \sum_{i=1}^{N} \mathcal{L}[\sigma^-_i]\rho \nonumber \\
&&  + \eta\sum_{i=1}^{N}\mathcal{L}[\sigma^+_i]\rho + \chi\sum_{i=1}^{N}\mathcal{L}[\sigma^z_i]\rho. \label{E1}
\end{eqnarray}
The system Hamiltonian is $H= \omega_c a^{+}a + \omega_a\sum_{i=1}^{N}\sigma^z_i + g\sum_{i} (a^+ \sigma^-_i + a \sigma^+_i) $ with photon creation $a^+$ and annihilation operator $a$ and the Pauli operators $\sigma^z_i,\sigma^-_i,\sigma^+_i$ of the $i^{\rm th}$ atom. Dissipation is accounted for by Lindblad terms with the superoperator $\mathcal{L}[c]\rho= c \rho c^{+} - \frac{1}{2}\{c^{+}c,\rho\}$, employed for different operators $c = a, \sigma_i^-,\ \sigma_i^+$ and $\sigma_i^z$ with appropriate rate factors.

Since the system is incoherently pumped, the exact solution of Eq.(\ref{E1}) does not produce a steady-state with a finite coherent field amplitude. While a symmetry breaking mean field ansatz may yield useful results, we expect a second-order mean-field theory without the symmetry-breaking assumption to offer solutions with a wider range of applicability. As in~\cite{PhysRevLett.102.163601} we obtain from the master equation (\ref{E1}) an equation for the mean intra-cavity photon number $\lan a^+ a \ran$, which couples to atom-photon correlations $\lan a \sigma^+_i \ran$ (identical for all the atoms). These correlations in turn depend on atom-atom correlation $\lan \sigma^+_i \sigma^-_j \ran$ (identical for any atom pair $i \neq j$) and the atomic population inversion $\lan \sigma^z_i \ran$. Eventually, higher order correlations like $\lan\sigma^z_i a^{+} a \ran$ appear, and to close the hierarchy of equations at second order, we utilize third-order cumulant expansion to approximate such higher order terms with products of non-vanishing lower order quantities such as $\lan\sigma^z_i \ran \lan a^{+} a \ran$, see Appendix \ref{App0}. Notice that the separated photonic and atomic coherence do not exist since  $\lan a\ran$ and $\lan \sigma^-_i \ran$ vanish, while they  have well defined relative phases through non-vanishing $\lan a^+ \sigma_i^- \ran$ and show phase stability over time via second-order temporal correlations, such as $\lan a^+(t) a(t+\tau) \ran \neq 0 $.

\begin{figure}
\centering
\includegraphics[scale=0.23]{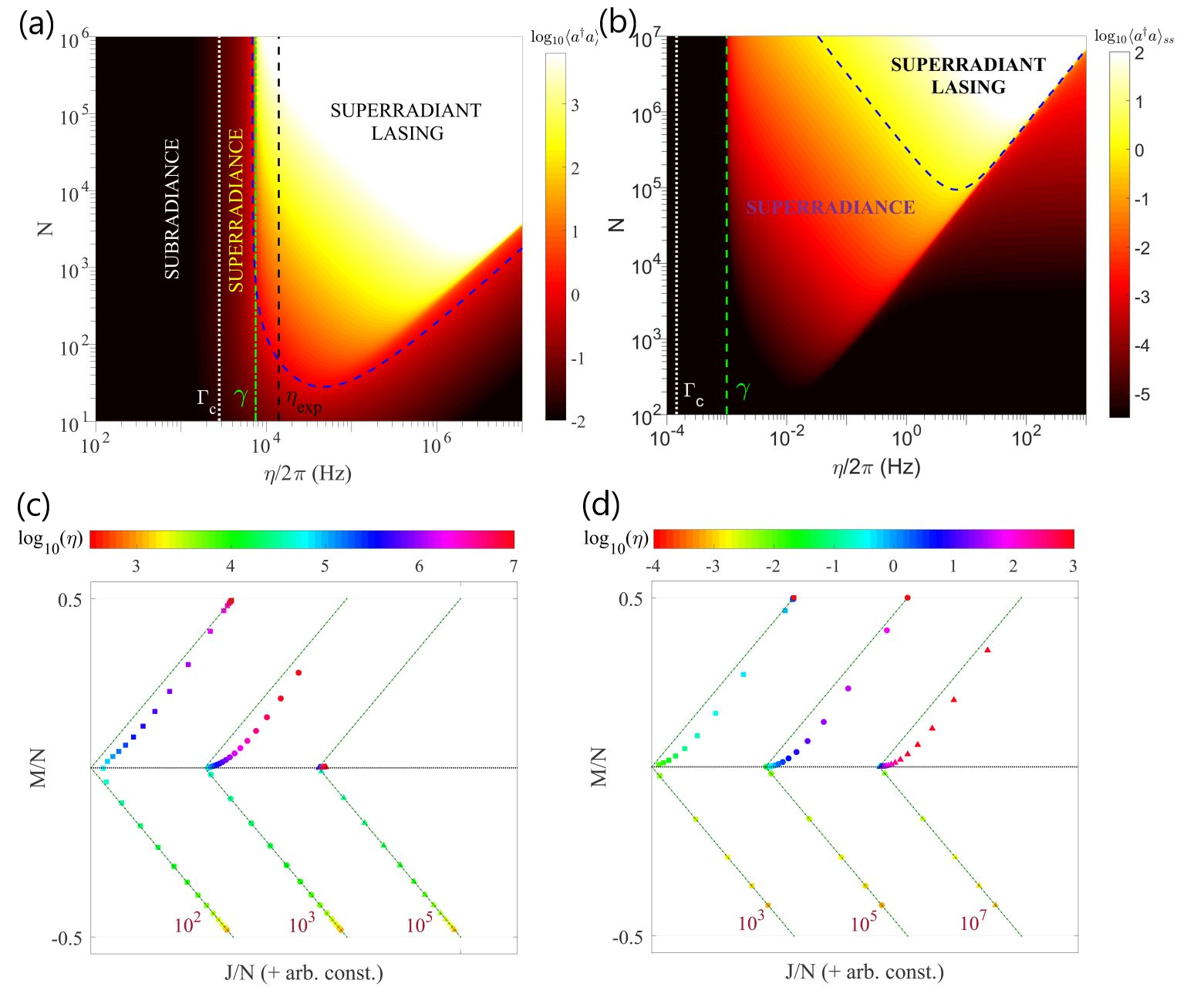}
\caption{ Intra-cavity photon number (a,b) and $J/N,M/N$ numbers (c,d)  from the systems using the ${}^3 P_1 \to {}^1 S_0$ transition of strontium-88 atoms (a,c) and the ${}^3 P_0 \to {}^1 S_0$ transition of strontium-87 atoms (b,d) for different $N$ and  $\eta$. In panel (a) and (b), the white dashed and black dash-dotted line indicate the Purcell rate $\Gamma_c= 2\pi \times 2.81$ kHz and $2\pi \times 0.15$ mHz, the atomic decay rate $\gamma= 2\pi \times 7.5$ kHz and $2\pi$ mHz, respectively, while the green dashed vertical line indicates  the pumping $\eta_{exp}= 2\pi \times 23.87$ kHz reported in~\cite{PhysRevX.6.011025}. The blue dashed line indicates single intra-cavity photon. In the panels (c,d) the results for different $N$ are shifted horizontally and the scaled $J/N$ and $M/N$ attain values in $[0,0.5]$ and $[-0.5,0.5]$, respectively. The colors indicate the varying pumping rates and the thin dotted green lines indicate the boundaries of the Dicke states.
}
\label{F2}
\end{figure}

\section{Steady-State Radiation \label{Radiation}}
The upper panels of Fig.~\ref{F2} show the steady-state intra-cavity photon number as function of the incoherent pumping strength $\eta$ and the number of atoms $N$ for the two strontium transitions. In  Fig.~\ref{F2} (a) we see that the photon number is very small if $\eta$ is weaker than the Purcell enhanced atomic decay rate $\Gamma_c = 4g^2/\kappa \approx 2\pi \times 2.81 $ kHz, which is identified as a subradiant regime ~\cite{PhysRevLett.102.163601}. When $\eta$ overcomes  $\Gamma_c$, the photon number increases but is still below unity, which is identified as the superradiant regime. Notice that single photons are indicated with the blue dashed contours. When $\eta$ overcomes the atomic decay $\gamma$, the photon number increases dramatically, which is then the superradiant lasing regime. For even larger $\eta$ the intra-cavity photon number decreases again. The range of $\eta$ leading to a large photon number becomes wider with increasing atom number. With the pumping reported in the experiment, we can achieve $5\times10^3$ cavity photons with $10^5$ atoms. In Fig.\ref{F2} (b) for the very narrow optical clock transition we observe similar results as for few hundred atoms on the wider transition in Fig.\ref{F2} (a). However, since the isolated and Purcell enhanced atomic decay rates are about six orders of magnitude smaller, the pumping rates considered here are also similarly smaller. 

This behavior, which was also observed in the generic analysis in ~\cite{TieriHolland}, is due to the collective atomic response to the pumping and interaction mechanisms. To illustrate this, we recall the symmetric collective states $\ket{J,M,\alpha}$ introduced by Dicke~\cite{PhysRev.93.99} with the quantum numbers $J$ and $M$ representing the eigenvalues of $\vec{J}^2$ and $J_z$. Here, $\vec{J} = \frac{1}{2}\sum_i \vec{\sigma}_i$ denotes the collective spin of the ensemble of effective spin-1/2 atoms, and $\alpha$ is an auxiliary quantum number to distinguish states with degenerate spin eigenvalues. Since all the atoms are identical, the states with different $\alpha$ have equal population and it is sufficient to utilize the symbol $\ket{J,M}$ to represent any of them. The Dicke state basis is equivalent in complexity to the SU(4) method~\cite{PhysRevA.87.062101}, and for small atom numbers it is useful for density matrix calculations~\cite{PhysRevA.78.052101}. Due to absence of coherences it may be also used for highly efficient stochastic wave function simulations of many atoms in a bad cavity~\cite{arXiv:1806.02156v1}. The superradiant coupling to the cavity mode and the ensuing emission are associated with the collective spin lowering operator $J^- =\sum_i \sigma^-_i$, which acts on the Dicke states by the familiar expression
\begin{equation} \label{eq:Dicke_lowering}
J^-|J,M\rangle = \sqrt{(J-M+1)(J+M)}|J,M-1\rangle.
\end{equation}
The collective transient superradiant decay of an initially excited atomic ensemble follows a progression of states $\ket{J,M}$ in the $J=N/2$ manifold of symmetric pure states. The initial and final transition amplitudes are collectively amplified by $\propto \sqrt{N}$ while during passage of states around $M\simeq 0$, the transition amplitude attains (superradiant) values $\propto N$. Due to atomic decay and pumping, our system will also explore symmetric states with $J$ less than $N/2$~\cite{PhysRevA.78.052101}, but the expression (\ref{eq:Dicke_lowering}) also yields the (reduced) transition amplitudes in these cases. Hence, it is useful to characterize the collective dynamics with the Dicke quantum numbers rather than the second order moments. To this end, we evaluate $M=\langle J_z\rangle = \frac{1}{2}\sum_i \langle \sigma_i^z \rangle$ and define $J$ as the square root of the expectation value of the operator $\vec{J}^2$, which is given by atomic second moments through $\langle \vec{J}^2 \rangle=\frac{3}{4}N + \frac{1}{4}\sum_{j\neq k} (\langle\sigma_j^- \sigma_k^+\rangle + \langle\sigma^z_j \sigma^z_k\rangle)$.

The lower panels of Fig.\ref{F2} show the mean Dicke quantum numbers for the steady state of the atomic systems accompanying the mean photon number shown in the upper panels. With increasing number of atoms from left to right, the separate series of points in each panel represent solutions for increased pumping rate, as indicated by their colors.

It is interesting to see how the increased incoherent pumping of the atoms causes their excitation ($M/N$) to increase. At first this occurs at the expense of a decrease in $J/N$, and the atomic state follows the lower Dicke states with $M\sim -J$. For stronger pumping we embark into population inverted states with positive values of $M$. The pumping now necessarily drives the states towards larger $J$ values. These behaviors agree with the analysis of quantum jumps in Dicke states, see~\cite{arXiv:1806.02156v1} and Appendix \ref{AppB}. For the lower atom number the moderate pumping leads to atomic states with $M < J$. These state have a higher emission rate, cf. Eq.(\ref{eq:Dicke_lowering}), than states with $M\simeq J$ obtained under stronger pumping. With experimentally realistic pumping rates it is not possible to fully invert  the atomic population and hence we do not observe the subsequent saturation and decrease in photon number for larger number of atoms. The behavior is similar in Fig.~\ref{F2}(d) for the narrow clock transition, where the photon number is however reduced due to the much weaker coupling to the inverted atomic ensemble.  

The above analysis shows that the system always occupy the Dicke states near to the boundaries, which allows us to apply the Holstein-Primakoff approximation to analyze the involved physics. For the Dicke states of a given $J$ along the lower boundary,  we have $\sum_i \sigma_i^+ \approx \sqrt{2J} b^+$, $\sum_i \sigma_i^- \approx  \sqrt{2J} b$ and $\sum_i \sigma_i^z = - J \mathbb{1}$ with oscillator creation $b^+$ and annihilation operator $b$. Then, the atom-cavity mode coupling becomes $ \sqrt{2J}g(b^+ a + b a^+)$, which shares the excitation generated by other means between the atoms and photons. Notice that the states with $M=-J$ and $J<N/2$ represent atomic excitation located in non-radiative states. For the Dicke states of same $J$ along the upper boundary, we have  $\sum_i \sigma_i^+ \approx \sqrt{2J} b$, $\sum_i \sigma_i^- \approx  \sqrt{2J} b^+$ and $\sum_i \sigma_i^z = J \mathbb{1}$ and the atom-cavity mode coupling $ \sqrt{2J}g(b a + b^+ a^+)$. This coherent coupling allows simultaneous creation of excitation in both the (inverted) atomic and photonic oscillators and creates atom-light entanglement responsible for non-vanishing cavity field and strong linewidth narrowing (see below). However, this coherent interaction dominates only if the collective coupling exceeds the cavity loss rate, i.e.  $\sqrt{N} g > \kappa$. This requires about $450$ strontium-88 atoms, which is well below the atoms achievable in the experiment. For strontium-87  it requires about $4.4\times 10^9$  atoms, which is about four orders of magnitude larger than achievable in current experiment. 

\section{Steady-State Spectrum \label{Spectrum}}
Superradiant lasing offers the possibility to obtain a light source with a linewidth set by the atomic lifetime rather than by the optical cavity. The two strontium transitions would thus lead to linewidth in the kHz and mHz regime, respectively. To resolve the linewidth of the field emitted by the cavity, we simulate its detection after passing through a filter cavity. Since the intensity transmitted is proportional to the photon number inside such a filter cavity and the field leaving the first cavity is proportional to its intracavity field, we may eliminate the traveling fields and obtain the spectrum by merely adding terms $-i[H_f,\rho]  + \mathcal{L}[\sqrt{\beta}f]\rho $ to the master equation (\ref{E1}). The Hamiltonian is $H_{f} =\omega_f f^+ f +  G(a^+f + a f^+) $ with filter cavity photon creation $f^+$ and annihilation operator $f$, mode frequency $\omega_f$, and the system-filter cavity coupling constant $G$. The photon loss in the filter cavity is described by the Lindblad term with a rate and hence cavity linewidth $\beta$.

The spectrum is obtained by calculating the filter cavity photon number $\lan f^+ f \ran$ for different values of $\omega_f$, and the line-width $\Delta \nu$ is extracted by fitting the spectrum with a Lorentzian function. To disregard reflection from the filter cavity into the main system and to properly resolve the spectrum, we require $G$ to be very small and $\beta$ to be smaller than $\Delta \nu$ for steady-state sources. This procedure to extract the emission spectrum is formally equivalent to the quantum regression theorem. The current approach can be straightforwardly implemented within the second order mean field theory where we merely include second moments involving the field observables of the filter cavity, see Appendix \ref{AppA}.

\begin{figure}
\centering
\includegraphics[scale=0.14]{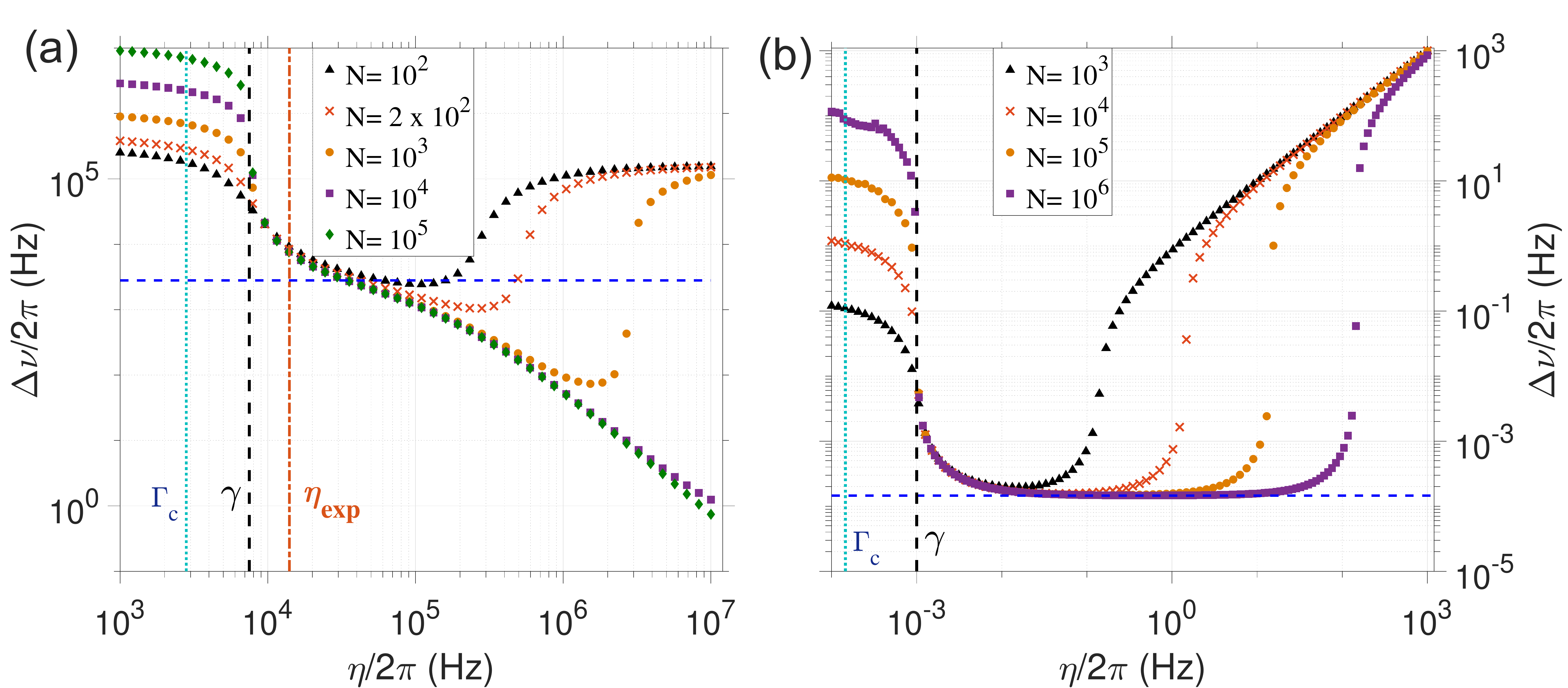}
\center
\includegraphics[scale=0.28]{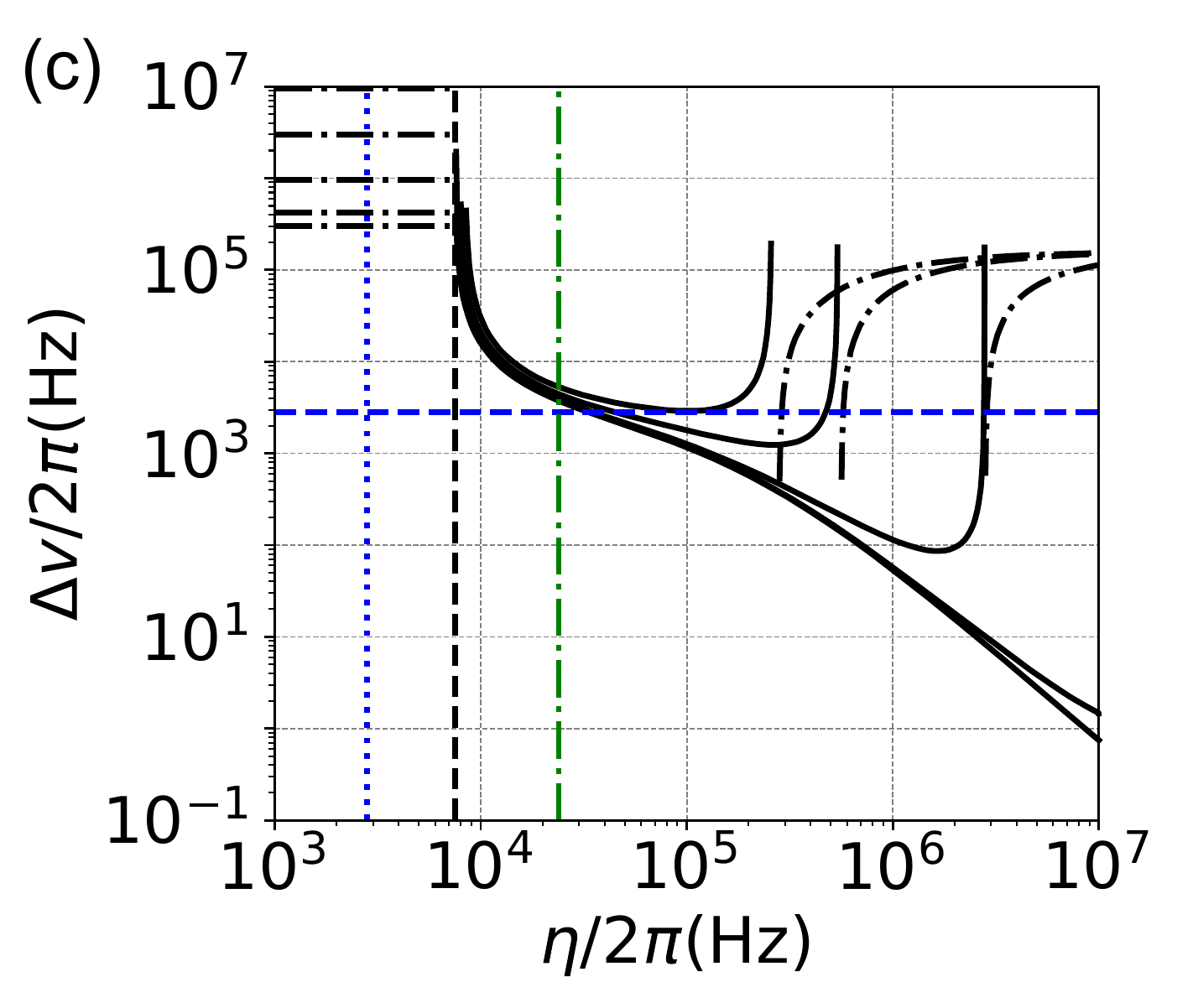}
\includegraphics[scale=0.28]{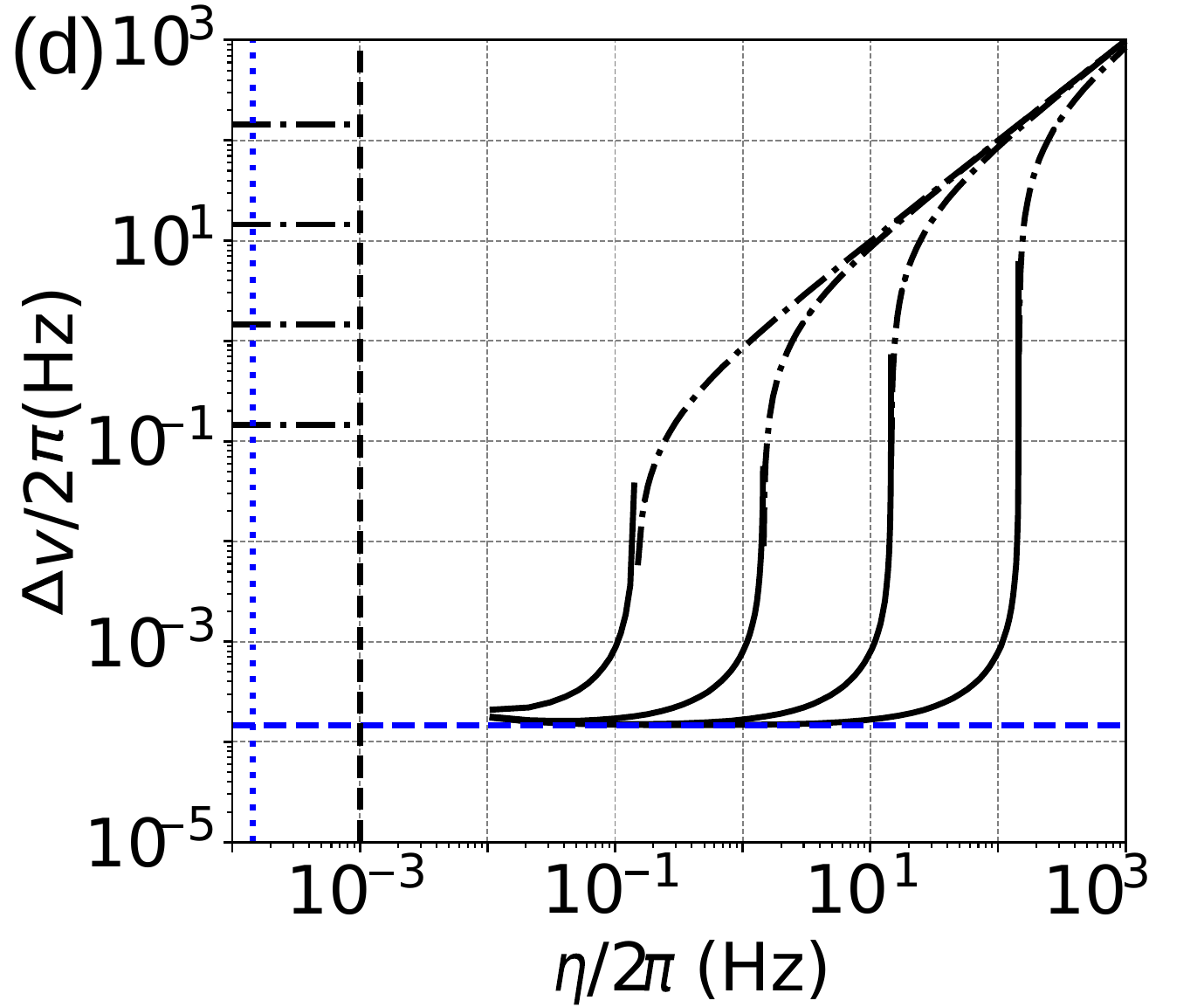}

\caption{Line-width $\Delta \nu$ of emission spectra from cavity coupled to the strontium-88 atoms through the ${}^3 P_1 \to {}^1 S_0$ transition (a), and the strontium-87 atoms through the ${}^3 P_0 \to {}^1 S_0$ clock transition (b). The discrete symbols show $\Delta \nu$  determined by our second-order mean field equations as function of incoherent pumping strength $\eta$ for different numbers of atoms $N$. The blue dashed horizontal lines indicate the minimal $\Delta \nu$ of the superradiant laser. The black and red vertical lines indicate the single atom decay rate $\gamma$ and the pumping rate $\eta_{\rm exp}$ achieved in experiments~\cite{PhysRevX.6.011025}. Panels (c,d) show corresponding analytical values for $\Delta \nu$. The expression (\ref{line-width}) predicted with mean field and laser diffusion theory~\cite{TieriHolland} (solid curves) works well for moderate pumping, while $2\sqrt{N}g$ and $N\Gamma_c$ [the left dot-dashed lines in the panel (c) and (d) respectively] and $(\Gamma\kappa - 4Ng^2)/(\Gamma +\kappa)$ (the dot-dashed lines on the right), resulted from Eq. (\ref{delta-nu}),  describe the weak and strong pumping better, respectively.}
\label{F3}
\end{figure}

Fig.~\ref{F3} shows the line-width $\Delta \nu$ of the steady-state spectra as function of pumping rate $\eta$  for the cavity coupled to the strontium-88 atoms via the ${}^3 P_1 \to {}^1 S_0$ transition, see panels (a,c), and to the strontium-87 atoms via the ${}^3 P_0 \to {}^1 S_0$ optical clock transition, see panels (b,d). The different symbols indicate different atom numbers $N$. The panel (a) shows that  for $N=10^2$ atoms (the black triangles) $\Delta \nu$ is constant for small $\eta$, and reduces slightly once $\eta$ overcomes the Purcell rate $\Gamma_c$ and drops significantly when $\eta$ exceeds the atomic decay rate $\gamma$. For much stronger driving $\Delta \nu$ increases again.

For the systems with more atoms, e.g. $N=2\times10^2,10^3,10^4,10^5$, $\Delta \nu$ attains larger values for weak pumping while it gets smaller for strong pumping. Once the system enters the superradiant lasing regime, the line-width approaches the Purcell rate $\Gamma_c$ (the blue dashed line).  For more atoms, however, $\Delta \nu$  decreases further. With $10^5$ atoms, as available in the experiment, the line-width may even reach the sub-hertz regime with the demanding pumping of $10$ MHz. 

This line-width narrowing may be related to atomic spin and photon entanglement (squeezing) resulted from the parametric coupling $ \sqrt{2J}g(b a + b^+ a^+)$ of the photonic and atomic oscillators in Holstein-Primakoff approximation.  With the maximal pumping rate $\eta_{\rm exp}$ achieved in the experiment~\cite{PhysRevX.6.011025}  (the orange dot-dashed vertical line), our theory yields a line-width, $\Delta \nu = 2\pi \times 6$ kHz, which agrees well with the experiment. 

In Fig.\ref{F3} (b) we study the system of strontium-87 atoms with the same cavity parameters as used for strontium-88 atoms. Due to the much weaker atomic dipole, the atomic decay rate and the coupling to the cavity field are much weaker, and the dynamical regime explored is different for the two systems. We find similar trends as in Fig.\ref{F3} (a) but $\Delta \nu$  is mainly determined by the atomic coherence due to the cavity loss being typically faster than the coherent excitation by the atoms. Without the stimulated emission and the coherence in the optical field contributing to the spectrum narrowing, the minimal line-width thus approaches the Purcell atomic decay rate $\Gamma_c= 2\pi \times 0.15$ mHz (the blue dashed line). We emphasize that to obtain the steady-state properties predicted here the atom loss occurring in  current experiments shall be avoided or compensated ~\cite{arXiv:1806.02156v1}.

It is possible to obtain good analytical expressions for the linewidths shown in Fig.\ref{F3} (a,b). Indeed,  using first order mean field and phase diffusion theory, originally developed for conventional laser~\cite{scully_zubairy_1997}, Tieri et. al.~\cite{TieriHolland} have achieved analytical expressions for the spectrum line-width in the superradiant lasing regime
 \begin{equation}
\Delta \nu = \frac{1}{2} \frac{C+\Gamma}{C d_0 -\Gamma} \frac{\Gamma}{\omega + \gamma} \frac{4g^2\kappa}{(\kappa + \Gamma)^2}
\label{line-width}
\end{equation}
with $C=N\Gamma_c$, $\Gamma = \eta  + \gamma + 2\chi$ and $d_0 = (\eta - \gamma)/(\eta + \gamma)$.
The calculations with this expression are shown as the solid curves in Fig.\ref{F3} (c,d), and agree with the exact numerical calculations for intermediate pumping strengths. For the strontium-88 transition, this agreement applies to the superradiant lasing regime with many photons, while, surprisingly, it also applies for the narrow strontium-87 transition in the superradiance regime where less than one photon is excited (for $N=10^3,10^4,10^5$ atoms). The validity of the mean field theory in this superradiant regime reflects that in the absence of a cavity mean-field the phase of the emitted field is determined by the stable correlation present in the excited atoms. 

The mean field expression (\ref{line-width}), however, fails for weak and strong pumping.  To address these regimes we consider the second order mean-field theory with steady-state solution for the filter cavity photon number
$ \lan f^{+}f\ran \propto - 2G {\rm Im} [\tilde{\omega}_2  \lan a^+  a \ran /(\tilde{\omega}_1 \tilde{\omega}_2 + N g^2 \lan \sigma^z_i \ran)]
- 2G {\rm Im} [Ng \lan a \sigma^+_i \ran /(\tilde{\omega}_1 \tilde{\omega}_2 + N g^2 \lan \sigma^z_i \ran]$ with $\tilde{\omega}_{1}= \omega_f -\omega_c +i (\beta + \kappa)/2$ and $\tilde{\omega}_2= \omega_f-\omega_a + i [(\beta + \gamma + \eta)/2 + \chi]$. The system observables $ \lan a^+  a \ran $ and $ \lan a \sigma^+_i \ran$ do not change under variation of  $\omega_f$, while their pre-factors attain the form of Lorentzian functions.  With this we obtain four candidate expressions for $\Delta \nu$ and identify the relevant one as
\begin{equation}
\Delta \nu = \frac{\Gamma+\kappa}{2} \left [\sqrt{1+4\frac{\Gamma/\kappa  - M  2\Gamma_c/\kappa   }{(\Gamma/\kappa+1)^2}} -1\right ].
\label{delta-nu}
\end{equation}
Note that the linewidth relies on the Dicke quantum number $M$ shown in Fig. \ref{F2} (c,d). 

For weak pumping with $\eta \approx 0$ and $\Gamma \approx 0$, we have $M=-J=-N/2$ and can approximate Eq.(\ref{delta-nu}) by the collective Purcell decay rate $N \Gamma_c$ if $2\sqrt{N}g \ll \kappa$, and by the collective Rabi frequency $2\sqrt{N}g$ if $2\sqrt{N}g \gg \kappa$. The former and latter expression fit the regime studied for strontium-87 and -88 atoms, respectively, and are represented by the dot-dashed lines in the left hand side of Fig.\ref{F3} (c,d). For $4Ng^2/\kappa <\Gamma \approx \eta \ll \kappa$  we have $M=J=N/2$ and can approximate Eq.(\ref{delta-nu})  by  $(\Gamma - N\Gamma_c)/(\Gamma/\kappa +1)$. This expression is shown by the dot-dashed lines in the right hand side  of Fig.\ref{F3} (c,d), and agrees well  with the numerical calculations shown in Fig.\ref{F3} (a,b). For very strong pumping such that $\Gamma \gg N\Gamma_c$, the expression is further simplified and $\Delta\nu$  approaches $\kappa$, reflecting the filtering of the noisy atomic emission by the system cavity.

For given atom number $N$, the minimal spectrum line-width is achieved with the steady-states characterized by $J\approx 0.1 N$ and $M \approx 0.1 N$. These states involve superpositions of product states $\ket{J,M} \ket{n}$ with total $J+M+n$ excitations, where $\ket{n}$ are photon number states. These superposition states have zero value for $\langle a\rangle$, but non-zero values for $\langle a \sigma^+_i \rangle$ and $\langle \sigma^+_i  \sigma^-_j \rangle$. In addition, they also lead to non-vanishing $J,M,\langle a^+ a \rangle$ as calculated in Fig. \ref{F2}. A rather wide range of states with  $n \approx \langle a^+ a \rangle$ are populated for  strontium-88 atoms while a small range of states with $n \approx 1$  are populated for strontium-87 atoms. This difference causes the different behavior of the spectrum linewidth as shown in Fig. \ref{F3}. 

\section{Conclusions}  In summary, we have studied lasing in the peculiar situation offered by ultra-cold strontium atoms trapped in an optical lattice inside an optical cavity, where coherence can be maintained in both the atoms and the cavity field. We showed that the system explores subradiance, superradiance and superradiant lasing regimes with increasing pumping rate. Using the long-lived optical clock transition of strontium-87 atoms, the atomic coherence dominates and our calculations confirm previous estimates of a millihertz optical emission line-width. Using the orders of magnitude broader ${}^3 P_1 \to {}^1 S_0$ transition of strontium-88 atoms, we remarkably obtain a line-width much smaller than the superradiant linewidth given by the single-atom Purcell rate. We associate this line-width narrowing with the coherence established between the field and the atoms under strong pumping of many atoms. To observe similar narrowing with the strontium-87 atoms it would request  orders of magnitude larger atom number as achieved in current experiment.

Our calculations are presented for experimental setups that are currently available in laboratories and may also be applied to other atoms, like calcium and ytterbium. The need for a significantly increased pumping rate and for methods to avoid or compensate atom loss are major challenges to test limiting cases of our theory. 

\begin{acknowledgments}
This work was supported by the Villum Foundation (Y. Zhang and K. M{\o}lmer) and
the European Union's Horizon 2020 research and innovation program
(No. 712721, NanoQtech, K. Debnath and K. M{\o}lmer) .
\end{acknowledgments}

\appendix
\section{Second-order Mean Field Equations \label{App0}}
In the main text, we outlined the procedure to solve the master equation (\ref{E1}) and we shall provide more details here. We start from the equation for the intra-cavity mean photon number 
\begin{equation}
\frac{\partial}{\partial t}  \lan a^{+} a \ran = -  2 g N {\rm Im} \lan a\sigma^+_i \ran  - \kappa\lan a^{+} a \ran,  \label{E3}
\end{equation}
which couples with the atom-photon correlation $\lan a \sigma^+_i \ran$. We assume all the atoms identical and use the symbol $\lan a \sigma^+_i \ran$ to represent the identical correlation with any atom. The factor $N$ comes from the summation over all the atoms.  In turn, the equation for this correlation is
\begin{eqnarray}
\frac{\partial}{\partial t} \lan a\sigma^+_i \ran &&= i \tilde{\omega}_{ac} \lan a\sigma^+\ran - i g \lan\sigma^z_i \ran\lan a^{+} a \ran \nonumber \\
 && - i g (N-1)\lan\sigma^+_i \sigma^-_j \ran - i(g/2)(1- \lan\sigma^z_i \ran). \label{E4}
\end{eqnarray}
Here, we have introduced the complex frequency $\tilde{\omega}_{ac} =  \omega_a - \omega_c  + i[ (\kappa + \gamma +  \eta) /2+\chi]$.  This correlation couples further with the atomic population inversion  $\lan \sigma^z_i \ran$ and the atom-atom correlation $\lan \sigma^+_i \sigma^-_j \ran$ ( $i \neq j$), which represents the correlation between any atom pair. The factor $N-1$ comes from the summation over all the atom pairs. The equations for the new quantities read

\begin{equation}
\frac{\partial}{\partial t} \lan\sigma^z_i\ran  = 4g {\rm Im} \lan a\sigma^+_i\ran - \gamma(1+\lan\sigma^z_i\ran) + \eta(1-\lan\sigma^z_i\ran), \label{E5}
\end{equation}
\begin{equation}
\frac{\partial}{\partial t} \lan\sigma^+_i \sigma^-_j \ran  = - 2 g\lan\sigma^z_i \ran {\rm}\lan a\sigma^+_i \ran - (\gamma + \eta + 2\chi )\lan\sigma^+_i \sigma^-_j \ran. \label{E6}
\end{equation}
During the derivation we encounter the expectation values of products of  three operators, e.g. $\lan \sigma^z_i  a^+  a \ran $, and the equations for them will again depend on the expectation values of products of four operators. To get closed equations only for the quantities introduced above, we apply third order cumulant expansion to replace them with the product of non-vanishing expectation values of two operators and of single operators, e.g. $\lan \sigma^z_i a^+ a\ran \to \lan \sigma^z_i \ran \lan a^+a\ran$. 

\begin{figure}[!h]
\centering
\includegraphics[scale=1.25]{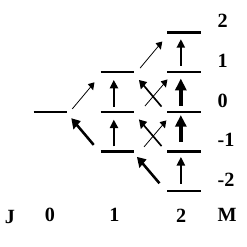}
\caption{ Incoherent transitions caused by atomic incoherent pumping are illustrated with arrows in Dicke states diagram for four atoms.  The arrow
thickness indicates the branching ratio of different transitions, see the text.
}
\label{FigApp}
\end{figure}
\section{Inconherent Transitions among Dicke States \label{AppB}}
In the main text, we connected the second-order moments with the Dicke states and then used the Dicke states to illustrate the system dynamics, see Fig. \ref{F2} (c,d). To understand this dynamics, the incoherent transitions caused by the atomic decay, dephasing and pumping have been identified in the Dicke states basis ~\cite{arXiv:1806.02156v1}. For this dynamics, it is however sufficient to consider the transitions related to the atomic incoherent pumping only. In the time interval $\delta t$  the pumping causes the transitions from the state $|J,M\rangle$ to $|J,M+1\rangle$ with the probability $\delta t \eta \frac{(2+N)(J-M)(J+M+1)}{4J(J+1)}$, to the state $|J - 1,M+1\rangle$ with the probability $\delta t \eta \frac{(N+2J+2)(J-M)(J-M-1)}{4J(2J+1)}$, and to the state $|J +1,M+1\rangle$ with the probability $\delta t \eta \frac{(N-2J)(J+M+1)(J+M+2)}{4(J+1)(2J+1)}$ ~\cite{arXiv:1806.02156v1}.  These transitions are illustrated as arrows between Dicke states for four atoms in Fig.  \ref{FigApp}.  We see that for the Dicke states along the lower boundary ($M=-J$) the jumps to the states with $J$ reduced by one have high probability than the jumps to other states. For the Dicke states along the upper boundary only the jumps to the states with $J$ increased by one are allowed. Together with the contribution from decay and dephasing  ~\cite{arXiv:1806.02156v1},  these processes determine the dynamics shown in Fig. \ref{F2} (c,d).

\section{Equations for Filter Cavity \label{AppA}}
To calculate the spectrum, we derive the equation for the mean photon number $\lan f^+ f \ran $ in the filter cavity:
\begin{equation}
\frac{\partial}{\partial t}\lan f^{+}f\ran = -2G{\rm Im}\lan af^{+}\ran - \beta\lan f^{+}f\ran
\label{A1}
\end{equation}
and find that it couples with the photon-photon correlation $\lan a f^+ \ran $. The equation for this correlation reads
\begin{equation}
\frac{\partial}{\partial t}\lan af^{+}\ran = i\tilde{\omega}_1 \lan af^{+}\ran + iG(\lan a^{+}a\ran + \lan f^{+}f\ran)  - igN\lan \sigma^-_i f^{+}\ran,
\label{A2}
\end{equation}
which further depends on the cross-cavity atom-photon correlation  $\lan \sigma^-_i f^{+}\ran$. Then, the equation for this new correlation reads
\begin{equation}
\frac{\partial}{\partial t}\lan \sigma^-_i f^{+}\ran =  i\tilde{\omega}_2\lan f\sigma^+_i \ran  + iG\lan \sigma^-_i a^{+}\ran + ig \lan\sigma^z_i \ran\lan af^{+}\ran.
\label{A3}
\end{equation}
Here, we have introduced $\tilde{\omega}_{1}= \omega_f -\omega_c +i (\beta + \kappa)/2$ and $\tilde{\omega}_2= \omega_f-\omega_a + i [(\beta + \gamma + \eta)/2 + \chi]$.
To close the equations for the quantities introduced so far, we approximate $\lan\sigma^z_i  af^{+}\ran$ by $\lan\sigma^z_i \ran\lan af^{+}\ran $.


%

\end{document}